\RequirePackage{fix-cm}
\documentclass[twocolumn]{svjour3}          
\smartqed  
\usepackage{graphicx}
\usepackage{epstopdf}
\usepackage{natbib}
\bibpunct{(}{)}{;}{a}{}{;}
\usepackage[utf8]{inputenc} 
\usepackage{xcolor}
\usepackage{ulem}
\usepackage{amsmath}
\let\dd=\partial

\let\*=\times

\usepackage{ulem}

\def\odier#1{{\textcolor{black}{ #1}}}    
\def\yvan#1{{\textcolor{black}{ #1}}}    

\begin{document}

\title{Mixing by internal waves quantified using combined PIV/PLIF technique 
}


\author{Dossmann Y. \and 
        Bourget B. \and
        Brouzet C. \and
        Dauxois T. \and
        Joubaud S. \and
        Odier P. 
}


\institute{
Univ Lyon, ENS de Lyon, Univ Claude Bernard, CNRS, Laboratoire de Physique, F-69342 Lyon, France 
 \\
P. Odier \at
          \email{podier@ens-lyon.fr}           
}

\date{\today~Received: date / Accepted: date}

\maketitle

\begin{abstract}

{We present a {novel} characterization of mixing ev\-ents associated with the propagation and overturning of internal waves, studied thanks to the simultaneous use of Particle Image Velocimetry (PIV) and Planar Laser Induced Fluorescence (PLIF) techniques. This combination of techniques had been developed earlier to provide an access to simultaneous velocity and density fields in  {two-layer} stratified flows  {with interfacial gravity waves}. {Here, for the first time, we show
how it is possible to implement it quantitatively}
 in the case of a continuously stratified fluid where internal waves propagate in the bulk. We explain in details how the calibration of the PLIF data is performed by an iterative procedure, and we describe the precise spatial and temporal synchronizations of the PIV and PLIF measurements. We then validate the whole procedure by characterizing the Triadic Resonance Instability (TRI) of an internal wave mode. 
{Very interestingly,} the combined technique is then applied to a {precise} measurement of the turbulent diffusivity $K_{\textrm{t}}$ associated with mixing events induced by an internal wave mode. Values up to $K_{\textrm{t}}=15~{\rm mm}^2\cdot {\rm s}^{-1}$ are reached when TRI is present
{(well above the noise of our measurement, typically $1~{\rm mm}^2\cdot {\rm s}^{-1}$)}, {unambiguously confirming} that TRI is a potential pathway to turbulent mixing in stratified flows.
{This work \odier{therefore provides a step}}}

\keywords{Internal waves \and  Mixing \and PLIF \and PIV \and PSI {\and TRI}}
\end{abstract}

\section{Introduction}
\label{intro}

Internal waves are ubiquitous in the ocean, due to the stratification in temperature and salinity. They are generated either from the interaction of tidal currents with submarine bathymetry~\citep{Garrett:ARFM:07}, or by wind stress at the ocean surface~\citep{Munk:DSR:98}. They can travel long distances, to reach places where they dissipate via various breaking mechanism \citep{Staquet:ARFM:02}. In the dissipation processes, they can produce mixing of the local stratification. These mixing processes are still scarcely understood, since in numerical simulations, they take place at sub-grid scales, {while in-situ measurements can only provide a discrete sampling of} oceanic regions (although breaking internal waves have been observed in the ocean, see~\citet{Lamb:ANFM:14} for a review). It is important to understand the mixing due to internal waves, since it could be one of the mechanisms to convert kinetic energy into potential energy, to maintain the ocean stratification against the tendency of settling more and more denser water at the bottom, via gravity currents~\citep{Kunze:Ocean:04,Wunsch:ARFM:04}.

Several experimental studies have dealt with the mixing induced by breaking internal waves. \citet{Thorpe:JFM:94:a} observed overturning waves using dye lines as qualitative tracers. \citet{Ivey:JFM:1989} measured the vertical mixing induced by the waves, using dye for qualitative visualization and conductivity probes for local quantitative measurement of the density profile evolution. The vertical evolution of a dye layer was also used to determine vertical diffusivity induced by the waves~\citep{Hebert:GRL:2003}. 

However, a direct characterization of mixing events would involve, for example, a detailed measurement of the vertical buoyancy flux, $g\,\langle\rho'w'\rangle/\overline\rho$, induced by the waves, where $\overline\rho$ is the average density, $g$ the gravity acceleration, $\rho'$ and $w'$ the fluctuating parts of the density and vertical velocity fields, while $\langle\cdot\rangle$ indicates ensemble- or time-averaging. In order to measure this quantity, a simultaneous quantitative measurement of the velocity and density fields in a continuous stratification is necessary. PIV is a very convenient technique to measure the velocity field. As to the density field, in general, in studies of internal waves, the synthetic Schlieren technique is used to measure the density gradient field, associated with wave propagation. Combined PIV-Schlieren measurements have permitted to describe the mechanical energy transported away by linear internal wave beams \citep{dossmann_simultaneous_2011}. 
However, as soon as mixing is involved, implying stochastic events and 3-D effects, the Schlieren technique becomes inaccurate, since it relies on the deterministic deviation of light beams by refractive index gradients. 

Another experimental technique used to measure quantitatively the density field is Planar Laser Induced Fluorescence (PLIF)~\citep{Karasso:EF:97}. Its principle is the following: the stratified fluid is seeded with fluorescent dye, proportionally to the local density difference. The fluid is then illuminated using a laser planar sheet, at the excitation wavelength of the dye. The light intensity emitted by the fluorescent dye is then proportional to the dye concentration. Taking pictures with a camera, the greyscale value in each pixel of the image allows for the determination of the local dye concentration. Providing that the diffusion coefficients of the fluorescent dye and the stratifying agent are similar (to avoid a double diffusive problem), the dye then serves as density marker. Using the same laser sheet as excitation light for PLIF and particle diffusive light for PIV, PLIF can usefully be combined with PIV to study jets~\citep{Hu:EF:00,Borg:EF:01,Feng:EF:07}, wakes~\citep{Hjertager:CJCE:03} and gravity currents~\citep{Odier:JFM:14}. {In the cases cited, a configuration of high optical contrast was obtained by using one dyed fluid (in general the jet or the current) and another undyed fluid (the ambient medium). The more difficult case of a continuous stratification characterized by a gradient of dye was tried by~\citet{Barret:PoF:91}, who {studied} grid turbulence in a stratified fluid, but, \odier{by the authors' own admission,} the PLIF data provided little quantitative measurement.} \odier{The principal reason for this is that in the former cases mentioned (gravity currents, jets), the variation of dye concentration extends over a few cm, which is the interface region between the jet or the current, and the ambient fluid, while in the case of~\citet{Barret:PoF:91}, the dye concentration gradients extends over a region of a few tens of cm. For this reason, the intensity variations associated to the motions studied are about 10 times smaller than in the former cases. }

This technique has also been used to {investigate} mixing induced by internal waves~\citep{Troy:ExpFl:2005,Hult:JGR:11a,Hult:JGR:11b}. However, {in these cases}, whether alone or combined with PIV, PLIF has only been used {to study waves at the interface of two fluids, one of which was dyed, providing again a highly contrasted configuration}. The aim of this article is to present an adaptation of the combined PIV/PLIF technique to study internal waves in a continuously stratified fluid, and to assess how information related to the mixing induced by the waves can be extracted by such a technique. {One of the challenges in this study is to be able to provide quantitative measurements of quantities like the buoyancy flux, in a continuously stratified environment where the PLIF signal/noise ratio will be much lower, \odier{about a factor 10, as explained in the previous paragraph,} than in the two-fluid configuration. {To the best of our knowledge, such quantitative extraction of information from a non-interfacial internal wave field using PLIF has never been performed before, so we consider that this work, in addition to presenting insights on how internal wave induced mixing relates to internal wave triadic instability, also provides \odier{a step} on the path to new possibilities of measurements for internal waves.}}\\

In section \ref{PLIF}, the experimental apparatus and PLIF technique are introduced, {as well as its coupling to PIV measurements}. The validation of the technique is exposed in section \ref{valid}, in particular in a configuration of {triadic resonance instability (TRI)}. The determination of eddy diffusivity induced by the {TRI} process is then presented in section \ref{pivlif}. Conclusions are drawn in section~\ref{conclu}. 


\section{Experimental techniques and data processing}
\label{PLIF}

\subsection{Fluorescence: emission and absorption.}
\label{basics}
Fluorescence is the capacity of an organic compound to absorb photons at a given wavelength $\lambda_{\rm abs}$, and reemit light at a different wavelength $\lambda_{\rm fluo}$. In the present experiments, rhodamine 6G was used, because its fluorescence is fairly independent of temperature, unlike rhodamine B. In addition, its excitation spectrum, peaking at 525 nm, is compatible with the laser we use (532 nm), unlike fluorescein (excitation peak at 490 nm). Its fluorescence emission peaks at 550 nm~\citep{Crimaldi2008}.

\begin{figure}
\includegraphics[width=0.5\textwidth]{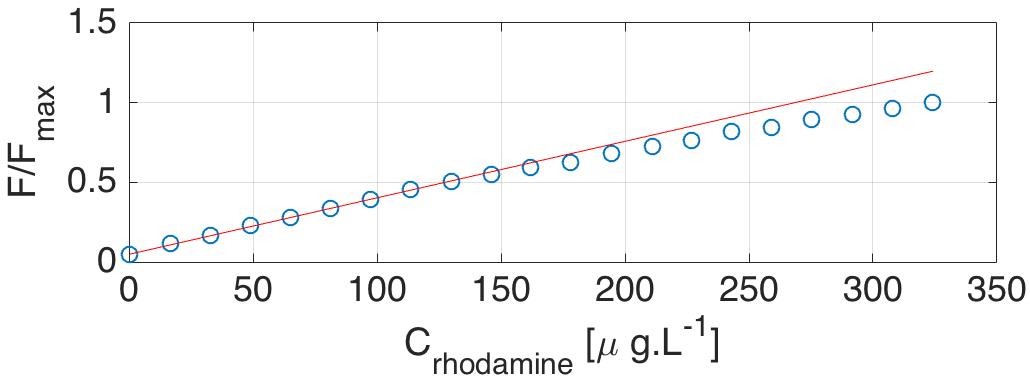}
\caption{Light intensity, $F$, emitted by a rhodamine 6G solution, normalized by its maximum value, versus its concentration. The line is a linear fit made on the range of concentration going from 0 to 120~$\mu$g$\cdot$L$^{-1}$}
\label{fig:linearite}       
\end{figure}

The light intensity $F$ emitted by a fluorescent solution can be expressed as~\citep{Patsayeva:99,Shan:EF:04}  
\begin{equation}
F\propto\frac{I}{1+I/I_{\textrm{sat}}}C \ ,
\label{eq:theorie_LIF}
\end{equation}
with $I$, the exciting light intensity, $I_{\textrm{sat}}$ the saturation intensity and $C$ the dye concentration. The linearity of the emitted intensity with the dye concentration was observed by~\citet{Shan:EF:04} up to concentrations of 48~$\mu$g$\cdot$L$^{-1}$. We extended the measurement range to higher values as shown in Fig.~\ref{fig:linearite}. {This measurement was made in an optically thin configuration, so that attenuation and reabsorption do not play a role.}
The linear behavior seems to prevail up to a dye concentration of 130~$\mu$g$\cdot$L$^{-1}$. Dye concentration below 100~$\mu$g$\cdot$L$^{-1}$ are used in the present experiments. Eq.~(\ref{eq:theorie_LIF}) also shows that the fluorescence emission is only linear with the excitation light intensity up to a certain point. The saturation intensity for rhodamine 6G is $5\times10^{9}$ W$\cdot$m$^{-2}$ \citep{Shan:EF:04}.


In the fluorescence process, energy is transferred from the excitation beam to the emitted light. The beam intensity thus decreases as it propagates through the fluorescent medium. This decrease is expressed by the Beer-Lambert law\begin{equation}
I(r)=I(r_0)\exp{\left(-\int^r_{r_0}\varepsilon C(s)\textrm ds\right)}\ ,
\label{absorptionintensite}
\end{equation}
giving the evolution of the light intensity for a beam propagating from point $r_0$ to point $r$, the distances being defined from the light source. The parameter $\varepsilon$ is the {mass} absorption coefficient of the dye and $C(s)$ its {mass} concentration along the optical path. Assuming linearity of the fluorescence process with the local excitation light intensity and with the local concentration, the fluorescent intensity along the optical path can be expressed as 
\begin{equation}
F(r)\propto I(r_0)\exp{\left(-\varepsilon\int^r_{r_0}C(s)\textrm ds\right)}C(r)\ .
\label{emissionintensite}
\end{equation}

\subsection{Experimental set-up}

\begin{figure*}
\includegraphics[width=0.98\textwidth]{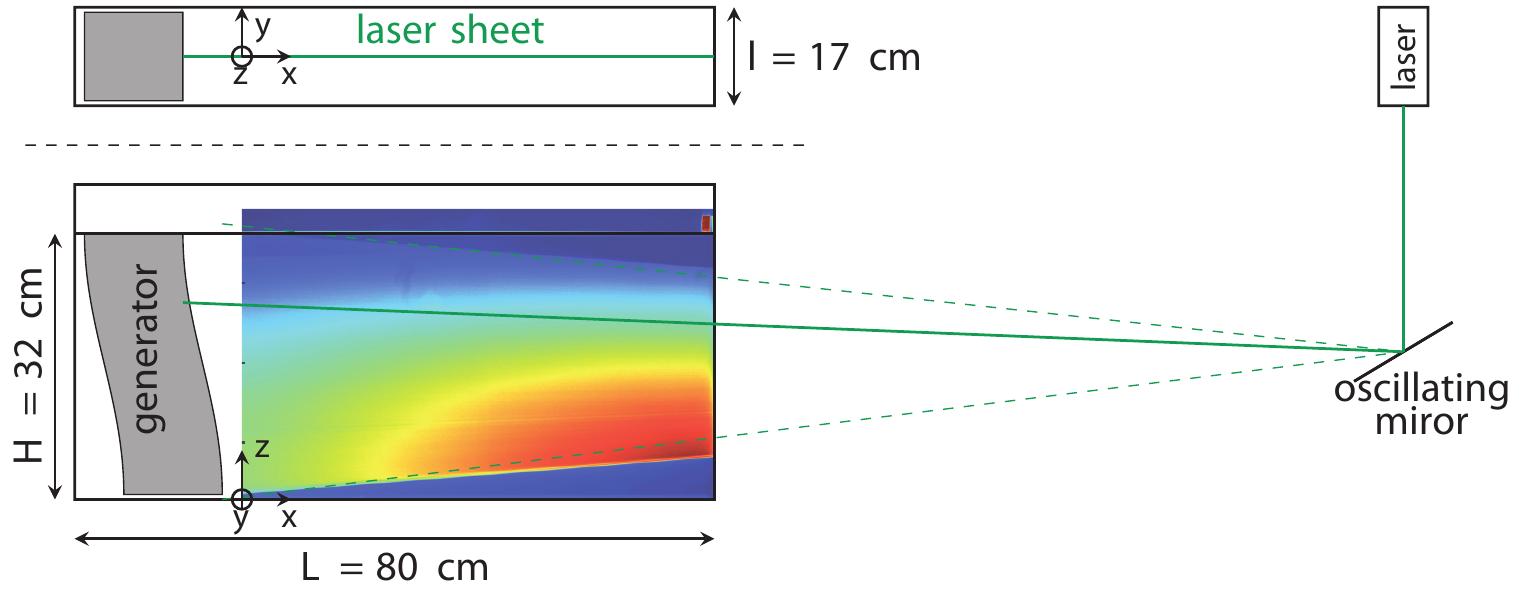}
\caption{The top panel presents the top view of the experimental set-up, while the bottom one shows a side view, with a raw image placed as a background. Note that the greyscale of the black and white camera has been replaced by a color scale for vizualisation purposes. The rest position of the front face of the wave generator is located at $x=$-5 cm. The calibration cell containing a fixed rhodamine concentration can be seen in the top right corner of the raw image. }
\label{fig:set-up}       
\end{figure*}

Experiments are conducted in a $80$~cm long, $17$~cm wide rectangular tank. It is filled to a height of $32$~cm with a linearly stratified fluid of density $\rho(z)$. Stratification is controlled via salinity. The buoyancy frequency $N=(-g\dd_z\rho/{\overline \rho})^{1/2}$ is about $1$~rad$\cdot$s$^{-1}$. A periodic internal wave is generated using a wave generator \citep{Gostiaux:EF:07,Mercier:JFM:10} placed vertically on a side of the tank. A mode-1 wave configuration propagating from left to right is forced (see schematics of the experimental set-up in Fig.~\ref{fig:set-up}). The generator forcing frequency is imposed using a Labview program.

Images are recorded with a CCD Allied Vision Pike camera, $14$-bits, $2452\times2054$ pixels, placed at $280$ cm from the tank. An optical filter (high-pass in wavelength with cut-off $550$ nm), is placed in front of the camera to allow only for the fluorescent light and to block the laser scattered light. The camera is mounted with a $35$~mm lens, producing a $40$ cm by $60$ cm field of view.

The PLIF excitation light is produced by a Laser Quantum Ti:Sapphire Opus laser of maximum output power $P=2$~W and wavelength $532$~nm, reflected on an oscillating mirror, creating a vertical light sheet, illuminating a vertical cut of the stratified fluid, in the middle of the tank width. The beam width is about $2$ mm, resulting in an excitation intensity of about $10^{6}$ W$\cdot$m$^{-2}$, much lower than the rhodamine 6G fluorescence saturation intensity $I_{\textrm{sat}}$, validating the linear approximation. 
Using an oscillating mirror allows for the production of a more homogeneous laser sheet, compared to the classical cylindrical lens method~{\citep{Crimaldi:EF:97}}. This makes the calibration procedure easier by preventing large calibration corrections due to illumination heterogeneities, {therefore yielding a uniform signal/noise ratio throughout the measurement domain}. The mirror oscillating frequency is adjusted so that an integer number of oscillations occur while the camera shutter is opened, providing a constant illumination for all images. Between two images, the mirror is non-moving, deflected outside the tank. This helps to minimize the photobleaching, which is the alteration of the fluorescent molecules by the laser light. 
{We checked, by measuring the time evolution of the fluorescence signal produced by a tank containing a {stationary} distribution of dye, that for the largest duration of our experiments (about 20 minutes), the photobleaching was negligible. It was indeed not measurable, compared to the 1 to 2\% fluctuations of the laser intensity.} 

{Sodium thiosulfate was added to the water, to neutralize the chlorine present in the tap water, preventing dye bleaching due to the chlorine.}

In general, a fluid stratified in density is also stratified in refractive index. Indeed, the stratifying agent, in our case salt, when mixed with water, increases its refractive index as well as its density. When mixing takes place, the index can then strongly vary locally, due to isopycnal overturns. Hence mixing events can induce transient (de)focalizing effects of incoming laser beams as previously observed by~\citet{Daviero:EF:01}. This effect  results in time dependent variations of local laser illumination, thus preventing proper calibration of the PLIF. For this reason, a refractive index matching was used by~\citet{Daviero:EF:01}. To accomplish that, ethanol is used as a second stratifying agent, since when mixed with water, it decreases its density while increasing its refractive index. We performed precise measurements of density and refractive index as a function of the salt and ethanol concentrations, coherent with the measurements of \citet{Daviero:EF:01}. The typical refractive index jump between salt water and freshwater is $\Delta n=0.080$ before refractive index matching. When salt and ethanol are used in a proportion of mass concentration of 1 to 2.9, the refractive index is uniform throughout the tank, with maximal variations of $0.02 \Delta n$. We observed that this procedure seems efficient to restore the necessary quality of the image, up to concentrations of 80 g$\cdot$L$^{-1}$ for ethanol and 28 g$\cdot$L$^{-1}$ for salt. 


Practically, the stratified tank is prepared using the standard two-bucket method~\citep{Fortuin:JPS:60,Oster:CR:63}, where the first bucket is filled with salt water at a chosen concentration, corresponding to the desired stratification, and the second bucket is filled with an ethanol solution {mixed with water} at {2.9} times the salt concentration of the first bucket. Rhodamine can then be added in either tank, depending on whether one wants the rhodamine concentration increasing or decreasing with depth. Note that the fluid viscosity can experience maximum changes of $20\%$ through depth due to the added ethanol. However we expect the flow to remain in the same regime as the Reynolds number remains in a narrow range (between 1000 and 1500 for a mode-1 wave of the amplitude considered here) for a fixed forcing frequency and this change should not have a direct effect on our measurements.

\subsection{Image processing}

The image processing described here is an adaptation to the case of continuous stratification of the calibration technique described in~\citet{Crimaldi2008} and references therein. In any experimental set-up using PLIF, the beam attenuation is not only due to the dye, but also, generally to a lesser extent, to all the other components of the fluid, namely water and stratifying agent. As a consequence, the attenuation expression in Eq.~(\ref{absorptionintensite}) must also take into account these other components. The unattenuated spatial intensity distribution of the beam, which we will denote $I_{\textrm{ua}}(x,z,t)$, (the subscript ``ua'' stands for ``unattenuated'') depends on the shape of the original beam and the optics forming the laser sheet. It decreases as $r^{-1}$ for the radial sheet used in the experiments presented in this paper. {Before any picture of the fluid was taken with the cameras, a spatial calibration was performed, using the PIV and PLIF camera images of a calibrated grid, in order to establish the correspondance between pixel location \odier{$(i,j)$} in an image and coordinates in the physical space \yvan{$(x(i,j),z(i,j))$}.} The greyscale value at a given point \yvan{$(x(i,j),z(i,j))$} in \yvan{the physical space} and at a given time $t$ can be expressed as

\yvan{\begin{multline}\label{eq:G}
G_{\textrm{cam}}(x,z,t)= G_{\textrm{b}}(x,z,t)\\
+\alpha I_{\textrm{ua}}(x,z,t) C_{\textrm{rhod}}(x,z,t) \exp{\left[-\beta(x,z,t)\right]}\,
\end{multline}
}

{\noindent where \yvan{$G_{\textrm{b}}(x,z,t)$} is the \yvan{camera} dark-res\-ponse\footnote{Experimentally, this term is computed by taking an image with the lens cap on, leaving only the camera noise as signal.} and $\alpha$ a proportionality coefficient characterizing the fluorescence efficiency, the camera sensitivity and the transfer function of the optical filter.} The light absorption term $\exp{\left[-\beta(x,z,t)\right]}$ in the last factor of Eq.~(\ref{eq:G}) is derived from Eq.~(\ref{absorptionintensite}), taking into account the absorption by all absorbents in the fluid. The absorption coefficients due to different absorbents are additive, so that in our case $\beta(x,y,t)$ can be expressed as

{\begin{equation}\label{eq:beta}
\begin{split}
\beta(x,z,t)=  \displaystyle\int_{r_{0}}^{r} & (\varepsilon_{\textrm{rhod}} C_{\textrm{rhod}}(s,t) +\varepsilon_{\textrm{salt}} C_{\textrm{salt}}(s,t) \\
& +\varepsilon_{\textrm{eth}} C_{\textrm{eth}}(s,t) +a_{\textrm{w}} )\textrm{d}s \ ,
\end{split}
\end{equation}}

\noindent {where $r$ is the path length at position $(x,y)$ along a ray path, $r_{0}$ the path length at the point of entry of the ray in the fluid, $\varepsilon_{\textrm{rhod}}$, $\varepsilon_{\textrm{salt}}$, $\varepsilon_{\textrm{eth}}$ the mass absorption coefficients, $C_{\textrm{rhod}}$, $C_{\textrm{salt}}$, $C_{\textrm{eth}}$ the mass concentrations and $a_{\textrm{w}}$ the absorption coefficient for water.} {In what follows, we will use the quantity $G=G_{\textrm{cam}}-G_{\textrm{b}}$ to eliminate the camera dark-response in the equations and keep only the greyscale level associated with fluorescence.}

In order to apply an absorption correction along the light path, the apparent position of the mirror needs to be determined (due to the variation of refractive index between air, plexiglas (PMMA) of the tank wall, and water, it is not the real position). To do that, an acquisition is performed at very low oscillating frequency of the mirror compared to the camera frame rate. Each image then gives the position of a ray at a given time. Extending these rays to the right, their common intersection is the virtual mirror location. This position is defined as the origin of a polar coordinate system which will be used for the absorption correction. 
This transformation {of the data into polar coordinate system} allows to compute the absorption correction simultaneously on each {ray path}, saving computational time.

In polar coordinates, the greyscale level associated with fluorescence can be rewritten as
\begin{equation}\label{eq:Gr}
G(r,\theta,t)=\alpha\, I_{\textrm{ua}}(r,\theta,t)\, \textrm{e}^{-\beta(r,\theta,t)}\  C_{\textrm{rhod}}(r,\theta,t) \ ,
\end{equation}
\noindent where $\beta$ is expressed by Eq.~(\ref{eq:beta}).

The laser intensity can vary slightly with time. We can safely assume that this variation is an overall variation, with no change in the spatial distribution (the typical time of these variations is much larger that the mirror oscillation period.) This assumption allows us to write $I_{\textrm{ua}}(r,\theta,t)=f(t)I_{\textrm{s}}(r,\theta)$. In order to take into account this time dependance, a small sample cell containing a rhodamine 6G solution of known concentration $C_{\textrm {ref}}$ is placed in the laser field (it can be seen in the top right corner of the image in Fig.~\ref{fig:set-up}). This cell being small, absorption can be neglected and the average greyscale value corresponding to the pixels in the cell can be expressed as $G_{\textrm{sol}}(t)=\alpha f(t)  I_{\textrm{cell}} C_{\textrm {ref}}$, where $ I_{\textrm{cell}}$ is the average laser intensity in the cell (the dark-response has also been subtracted to obtain $G_{\textrm{sol}}$). A renormalized greyscale value for the whole image can then be defined, canceling out the time dependance of the laser intensity
\begin{equation}  \label{eq:renormalisation}
\begin{split}
H(r,\theta,t) & =\frac{G(r,\theta,t)}{G_{\textrm{sol}}(t)}  \\
& = \frac{I_{\textrm{s}}(r,\theta)}{ I_{\textrm{cell}}} \frac{C_{\textrm{rhod}}(r,\theta,t)}{C_{\textrm {ref}}} \exp\left(-\beta(r,\theta,t)\right).
\end{split}
\end{equation}

In order to correct for absorption, the factor $I_{\textrm{s}}(r,\theta)$ needs to be estimated. To do that, a direct measurement of the initial stratification (before one generates waves) is used. This measurement is made by a conductivity probe that is slowly immersed in the stratified fluid, down to the bottom.\footnote{In order to take into account the presence of ethanol in the fluid, in addition to the salt, the conductivity probe must be calibrated with the mixture of salt and ethanol in the same proportions, using a density meter Anton Paar DMA35.} This measurement gives access to all the concentrations appearing in equation~(\ref{eq:beta}).

\begin{figure*}
\includegraphics[width=0.98\textwidth]{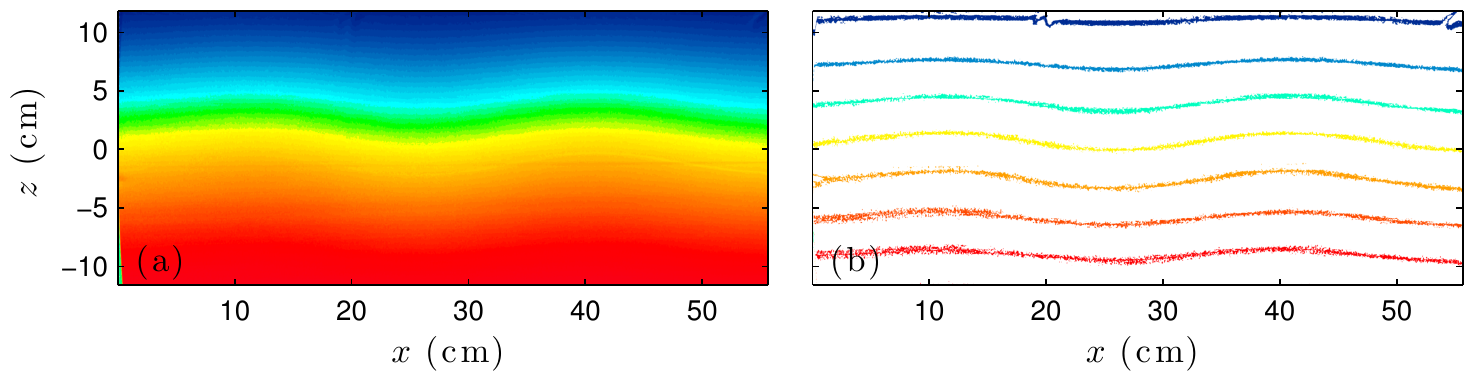}
\caption{Rhodamine concentration field $C_{\textrm{rhod}}$ when a mode-1 wave is propagating, obtained from Eq.~(\ref{eq:Crhod}). {\bf (a)} Full field. {\bf (b)} Selected isopycnals of the same field. Colors represent the concentration in rhodamine which goes from zero at the top (blue) to $40~\mu$g$\cdot$L$^{-1}$ at the bottom (red).}
\label{fig:corrigee2}       
\end{figure*}

{ The values of the mass absorption coefficients $\varepsilon$ for salt, etha\-nol and rhodamine 6G and of the absorption coefficient for water $a_{\textrm{w}}$ can be found in~\citet{Daviero:EF:01}. However, using literature values did not seem to optimize the absorption correction. In addition, these values rely on measurements using an argon laser, while we use a Ti:Sapphire laser, with a different wavelength. For this reason, we chose to adjust these values in order to optimize the correction. This was done by finding the values that provide an evolution along $r$ of the quantity $I_{\textrm{ua}}$ closest to the expected evolution, namely a $1/r$ decrease. We found $\varepsilon_{\textrm{rhod}}=5.6\cdot10^{-4}$~cm$^{-1}\cdot \mu$g$^{-1}\cdot$L, $\varepsilon_{\textrm{eth}}=3.5\cdot10^{-5}$ cm$^{-1}\cdot$g$^{-1}\cdot$L, $\varepsilon_{\textrm{salt}}=1.2\cdot10^{-4}$~cm$^{-1}\cdot$g$^{-1}\cdot$L and $a_{\textrm{w}}=3\cdot10^{-3}$ cm$^{-1}$. \odier{Keeping in mind that a different wavelength was used, making comparisons delicate, these values are close to the ones found in~\citet{Daviero:EF:01}, except for $\varepsilon_{\textrm{rhod}}$, which is about twice larger, and $\varepsilon_{\textrm{eth}}$, which is an order of magnitude smaller.}}

{It is interesting to note that with the concentrations used in this experiment, the attenuation due to water, salt and ethanol individually amount to about 10\% of the attenuation due to rhodamine 6G, which {requires} taking them into account.}

%
%
%
%
%

{Knowing the values of the concentrations and \linebreak absorption coefficients, the absorption fraction \linebreak $\exp\left(-\beta(r,\theta,0)\right)$ can be computed for the image at $t=0$ (before the waves start). The laser illumination field $I_{\textrm{s}}(r,\theta)$ can then be extracted  using equation (\ref{eq:renormalisation}) taken at $t=0$}
{\begin{equation}\label{eq:Is}
I_{\textrm{s}}(r,\theta) =\frac{H(r,\theta,0)I_{\textrm{cel}} C_{\textrm {ref}}}{C_{\textrm{rhod}}(r,\theta,0) \exp\left(-\beta({r},\theta,0)\right)}.
\end{equation}}

%
%


The dye concentration at time $t$ and position $(r,\theta)$ can then be expressed as
\begin{equation}
C_{\textrm{rhod}}(r,\theta,t)=\frac{H(r,\theta,t)I_{\textrm{cel}} C_{\textrm {ref}}}{I_{\textrm{s}}(r,\theta)}\exp\left(\beta(r,\theta,t)\right)
\end{equation}
where $I_{\textrm{s}}(r,\theta)$ is taken from Eq.~(\ref{eq:Is}).

However, the exponential factor $\exp\left(\beta(r,\theta,t)\right)$ now depends on the instantaneous concentrations along the ray path between the entry in the tank and the considered point $(r,\theta)$. For this reason, it can only be computed iteratively, using the dye concentration already computed {at the same time $t$ and at lower values of~$r$} \citep{Odier:JFM:14}. Assuming that there is no double diffusion, the concentrations of salt and ethanol, also needed to compute $\beta(r'<r,\theta,t)$, can be derived from the dye concentration.

The dye concentration at a given distance $r$ from the mirror can then be expressed iteratively
\begin{equation}\label{eq:Crhod}
\begin{split}
C_{\textrm{rhod}}(r+\textrm{d}r,\theta,t)= & C_{\textrm{rhod}}(r,\theta,t)\frac{H(r+\textrm{d}r,\theta,t)}{H(r,\theta,t)} \\
& \hskip -1.48 cm \times \frac{I_{\textrm{s}}(r,\theta)}{I_{\textrm{s}}(r+\textrm{d}r,\theta,t)}\exp\left(\frac{\partial \beta}{\partial r}(r,\theta,t)\textrm{d}r\right).
\end{split}
\end{equation}

This method uses the known dye concentration field at initial time, which is more accurate than the standard PLIF calibration method using a tank of uniform concentration prepared before or after the actual data runs.

{Figure~\ref{fig:corrigee2} shows an example of corrected PLIF image, when the wave generator is in motion, generating a mode-1 propagating horizontally. One can observe the displacement of the isopycnals, due to the wave. One can also observe that the mean position of the isopycnals is horizontal, attesting that the absorption correction was done accurately.} In case of an error in the model (absorption coefficients, concentrations), the isopycnals in the initial stratification loose their horizontality. This visualization also allows us an estimate of the noise introduced by the correction: in Fig.~\ref{fig:corrigee2}, the dye concentration varies from 0 to 40 $\mu$g$\cdot$L$^{-1}$ and the width of the isopycnals shown in Fig.~\ref{fig:corrigee2}b is 0.4 $\mu$g$\cdot$L$^{-1}$, which is $1\%$ of the vertical density variation. Since the stratification is linear and extends over a 32 cm depth, the width of the isopycnals should be 3.2 mm. In Fig.~\ref{fig:corrigee2}b, their actual width can be estimated to 4 mm, which is close to the expected value. This difference is due to the camera noise, as well as the noise introduced by the correction.

\subsection{Coupled PIV-PLIF device}


{In order to measure the velocity field,} the fluid is seeded with PIV particles (hollow glass spheres, density 1.1 kg$\cdot$m$^{-3}$, average diameter $8~\mu$m). The settling velocity of these particles, slightly more dense than the fluid, is negligible compared to the fluid velocity induced by the waves. Next to the PLIF camera (about $10$ cm apart), a second identical camera is placed, mounted with the same lens and a band-pass optical filter (530 $\pm$ 5 nm), which blocks the fluorescence light while passing the laser light, scattered by PIV particles. To avoid any acquisition problem, two computers are used to record the data, one for PIV camera and one for PLIF camera. The velocity fields are obtained by applying a PIV algorithm, based on the Fincham \& Delerce algorithm~\citep{Fincham:EF:00}, on the image pairs provided by the PIV camera. The PIV interrogation window size is 21$\times$21 pixels and the spatial resolution for the velocity field is $2.5$~mm, while the spatial resolution for the density field is 0.25 mm. If necessary, it can be coarse-grained to match the PIV resolution.

To compute velocity-density correlations, one must ensure that the velocity and density measurements are performed at the same time and at the same location. For time synchronization, the two cameras are triggered with the same signal issued from one computer. It consists in periodic square pulses controlled by a Labview program. The time interval between two pulses is $T_{\textrm{acq}}=1/6$~s, {which is adapted to the slow time evolution of the waves, but a faster frame rate could be used without problems, should one want to measure faster phenomena}. Hence, PLIF and PIV images are synchronized. 

The same signal is sent to an Agilent generator which {provides} the triangle signal controlling the oscillating mirror. 
About $60$ PIV and PLIF images are recorded per forcing period. When processing a couple of PIV images recorded at subsequent times $t_1$ and $t_2=t_1+T_{\textrm{acq}}$, the measured velocity field is actually averaged between $t_1$ and $t_2$. Hence $t_{1-2}\equiv(t_1+t_2){/2}$ is the actual measurement time for the velocity field. To account for the lag of $T_{\textrm{acq}}/2$ between velocity and density measurements, the average density $\rho(x,z,t_{1-2})=(\rho(x,z,t_{1})+\rho(x,z,t_{2})){/2}$ between $t_1$ and $t_2$ is computed. 
 

\begin{figure}
\includegraphics[width=0.5\textwidth]{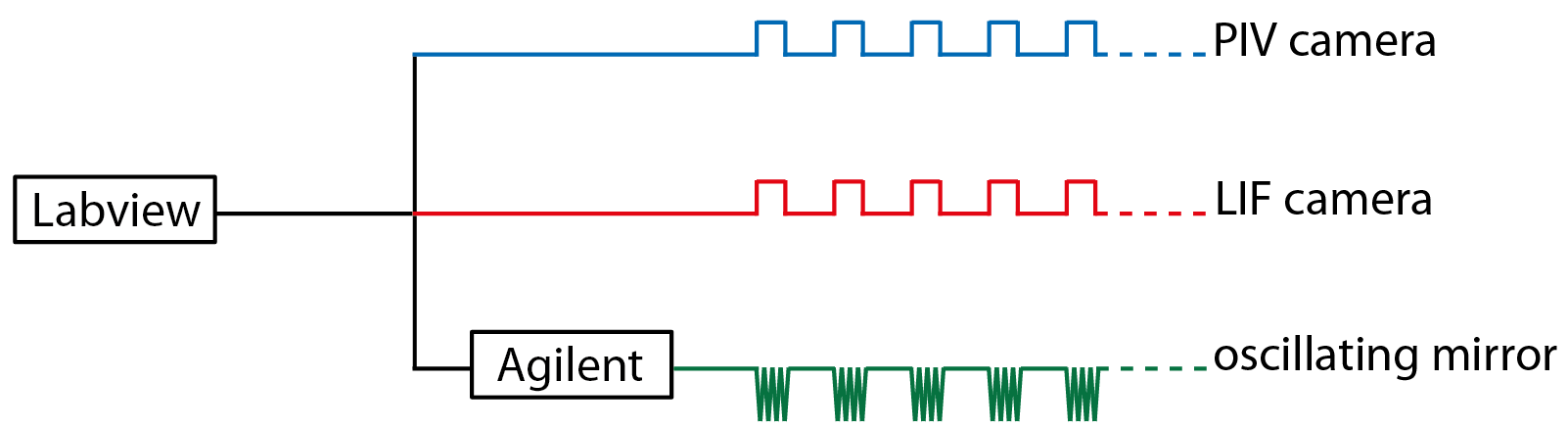}
\caption{Schematics of the set-up used for temporal synchronization between PIV and PLIF cameras.}
\label{fig:synchro}       
\end{figure}

In order to perform the spatial adjustment of the cameras, a calibration grid was placed in the field of view. Each camera provides an image of this grid. Comparing these images with the real space grid, two spatial transformations are defined: the first one transforms the PLIF image into the real space and the second one transforms the PIV image into the real space. This allows {us} to define corresponding grid points for the velocity and density measured fields.


\section{Validation of the procedure}
\label{valid}

\subsection{Wave detection through PIV/PLIF measurements}

\begin{figure}
\includegraphics[width=0.52\textwidth]{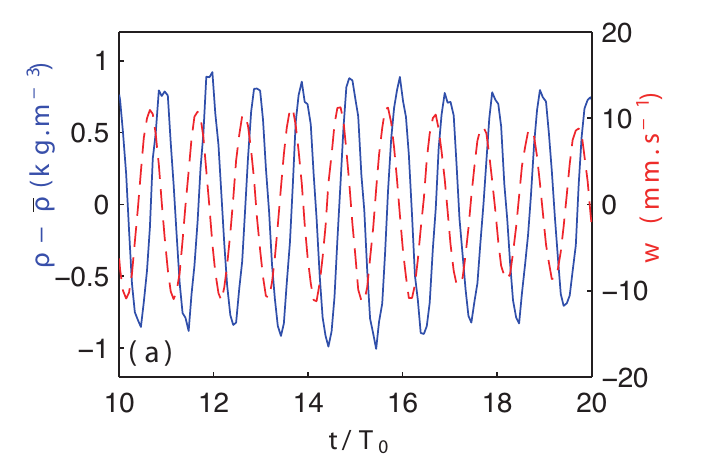}\\
\includegraphics[width=0.44\textwidth]{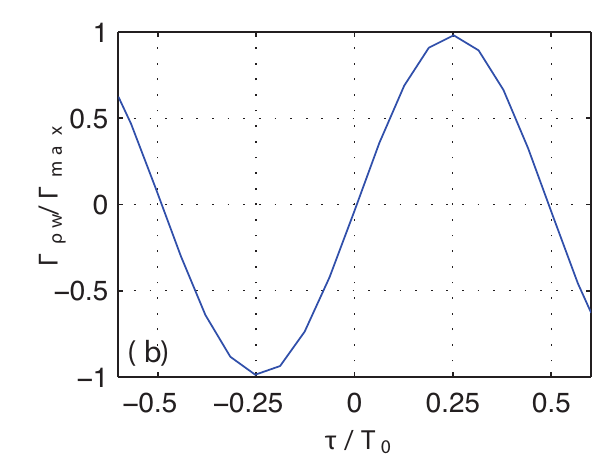}
\caption{{\bf (a)} The solid blue curve presents the density fluctuations, $\rho'=\rho-\overline\rho$ measured at the center of the field of view, vs time ($T_0=2\pi/\omega_0$ is the forcing period). Experimental parameters: $\omega_0/N=0.88$, mode-1 plate amplitude: $0.5$~cm. The red dashed curve corresponds to the vertical velocity of the fluid measured at the same point. {\bf (b)} Cross-correlation density/vertical velocity, $\Gamma_{\rho w}(\tau)$, as a function of non-dimensional delay, $\tau/T_{0}$.}
\label{fig:PIVLIF}       
\end{figure}

As an example of the results that can be obtained with this coupled technique, Fig.~\ref{fig:PIVLIF}a shows vertical velocity and density time series measured at a location chosen at the center of the field of view, once the spatial transformation has been applied to the two measured fields. For a plane wave or a vertical mode, the following relation can be derived between vertical velocity and density, from the mass conservation equation

\begin{equation}\label{w_rho}
w=-\frac{1}{\partial_z \overline\rho}\partial_t\rho .
\end{equation}

The time traces of vertical velocity and density must thus have a quarter-phase delay, which can be qualitatively observed in Fig.~\ref{fig:PIVLIF}a. To get a more precise information, one can compute the cross-correlation of the two signals, defined as

\begin{equation}
\Gamma_{\rho w}(\tau)=\int\rho(u)\,w^*(u-\tau)\textrm du.
\end{equation}

The result is shown in Fig.~\ref{fig:PIVLIF}b, where one can see that the maximum is obtained for a quarter of the wave period, confirming that the vertical velocity behaves as the derivative of the density (the minus sign in Eq.~(\ref{w_rho}) is cancelled out by the negative sign of $\partial_z \overline\rho$). The correct spatial and temporal adjustment between the two cameras is thus verified.

\subsection{Observation of { triadic resonance instability (TRI) }}

For a more quantitative test of the calibration procedure, we use the obtained density field to study the production of two secondary plane waves from a primary mode-1 wave. This is the result of a {triadic resonance} instability (TRI). \footnote{Note that the more commonly used acronym PSI (Parametric Subharmonic Instability) actually corresponds to a particular case of TRI in the case where viscosity is negligible; both unstable secondary waves then have a frequency equal to half of the forcing frequency. We prefer to use TRI to keep the generality.} This instability {will prove crucial in the mixing processes that we will assess in the next section.} It was studied in detail in~\citet{Joubaud:PoF:12} {and} \citet{Bourget:JFM:13} using the standard Schlieren technique to observe the waves. We try here to recover the results using the PLIF data, {as a test}. This instability is a resonant 3-wave interaction, which is inherent to internal waves. The two secondary waves fulfill with the primary wave a temporal resonance condition $\omega_0=\omega_1+\omega_2$, where $\omega_0$ is the primary wave frequency and $\omega_1$, $\omega_2$ the secondary waves frequencies, as well as a spatial resonance condition $\bf{k_0}=\bf{k_1}+\bf{k_2}$, where $\bf{k_0}$ is the wave vector of the primary wave and $\bf{k_1}$, $\bf{k_2}$ the wave vectors of the secondary waves. We will use the PLIF data to identify the secondary waves and measure their frequencies and wave vectors, for a mode-1 primary wave. 

\begin{figure}
\includegraphics[width=0.49\textwidth]{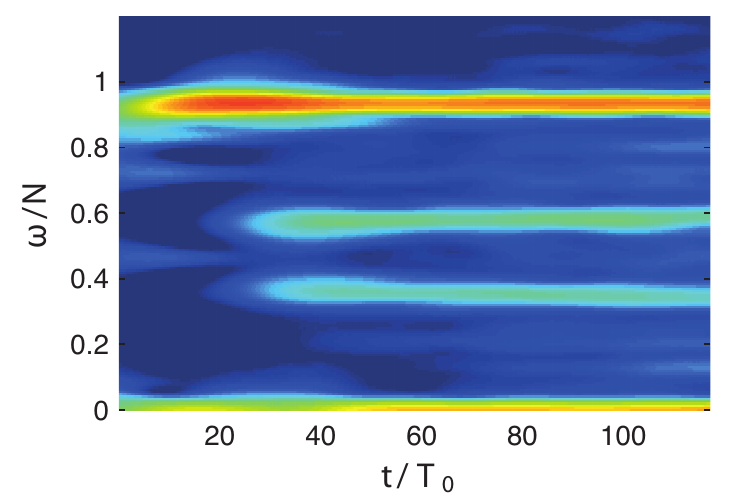}
\caption{Time frequency spectrum obtained from the density field. The primary wave is a mode-1 at normalized frequency $\omega_0/N=0.93$ ($N=0.9$ rad$\cdot$s$^{-1}$) and for a generator maximum plate displacement of $0.75$ cm.
}
\label{fig:time-frequency}       
\end{figure}

\begin{figure}
\centering
\includegraphics[width=0.38\textwidth]{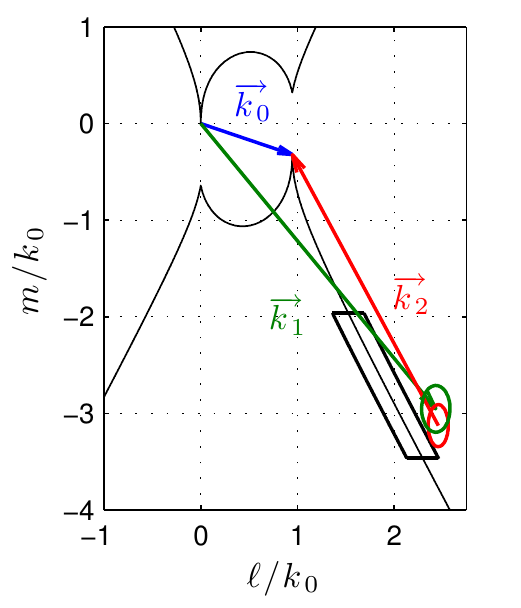}
\caption{In the $(\ell/k_0,m/k_0)$ plane (horizontal and vertical components of the wave vectors, normalized by the module of ${\textbf k_0}$), the three arrows are the  measurements of the three wave vectors. The dark solid line represents the theoretical resonance loci for the secondary wave vector ${\textbf k_1}$ for a given~${\textbf k_0}$. The rectangle represents the theoretical most unstable mode {while} the circles represent the measurement error on the wave vectors.}
\label{fig:vectors}       
\end{figure}

Figure~\ref{fig:time-frequency}a shows a time-frequency spectrum computed from the density field. One clearly sees the primary wave at frequency $\omega_0/N=0.93$ and after about 25 wave periods, two secondary frequencies appear: \linebreak $\omega_1/N=0.58$ and $\omega_2/N=0.34$. Their sum is equal to the primary wave frequency, verifying the temporal resonance condition. In order to measure the wave vectors, a temporal filter is applied around the 3 measured frequencies. Then, a 2D spatial spectrum is computed for each filtered signal, giving access\footnote{The field of view does not have enough vertical extension to allow a measurement of the vertical component of $\bf{k_0}$. However, from the design of the wave generator, we know that this component is equal to $\pi/H$, where $H$ is the generator height.} to $\bf{k_0}$, $\bf{k_1}$ and $\bf{k_2}$. The  vectors obtained are shown in Fig.~\ref{fig:vectors}, where one can check that they verify the spatial resonance condition. 

Resolving the system of equations formed by the temporal and spatial resonance conditions, together \linebreak with the dispersion relation for the 3 waves, one can compute in the $(\ell,m)$ plane the locus of the tip of vector $\bf{k_1}$ (see~\citet{Bourget:JFM:13}). It is represented as a solid black curve in Fig.~\ref{fig:vectors}. The measured value of $\bf{k_1}$ falls in the vicinity of this curve. Following again~\citet{Bourget:JFM:13}, by computing the growth rate of the instability for the experimental conditions, we estimate where in this curve vector $\bf{k_1}$ should fall. This position, including error estimates, is represented in Fig.~\ref{fig:vectors} by a rectangle. One can observe that within errors (those due to the measurement of the amplitude of the primary wave and those due to the measurement of $\bf{k_1}$), the experimental measurement of $\bf{k_1}$ is compatible with its theoretical expectation, validating the calibration procedure for the PLIF.


\section{Assessment of mixing processes}
\label{pivlif}

\begin{figure}
\includegraphics[width=0.5\textwidth]{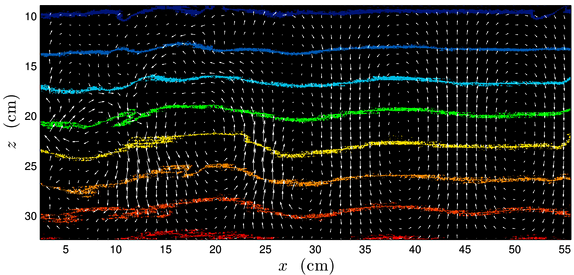}
\caption{Mixing event observed with the PIV/PLIF set-up. The arrows represent the velocity field, while the colored lines represent selected isopycnals. Experimental parameters: $\omega_0/N=0.92$, mode-1 plate amplitude: $0.75$~cm.}
\label{fig:mixing_event}       
\end{figure}

The initial goal in applying the coupled PIV/PLIF technique to an internal wave set-up was to be able to characterize the mixing induced by overturning waves. A qualitative picture of such an event can be observed in Fig.~\ref{fig:mixing_event}, showing the superposition of the velocity field measured by PIV (arrows) and of the density field, represented by selected isopycnal (colored lines). The velocity field shows the mode-1 wave rolls, superimposed with regions of more turbulent motion (on the left). The isopycnals illustrate local overturn events, coinciding with the turbulent regions of the velocity field. In these regions, the combination of these overturn events and the presence of turbulence suggests that mixing is likely to take place.

In order to get a more quantitative picture, the simultaneous measurement of density and velocity gives access to quantities based on the correlation between these two fields. For example, the buoyancy flux, defined as $(g/{\overline\rho})\langle\rho'w'\rangle$  where $\langle\cdot\rangle$ stands for the Reynolds average, represents the amount of buoyancy that is transported vertically by the velocity fluctuations, providing turbulent mixing. In oceanic simulations, it is often parameterized using the eddy diffusivity assumption, which assumes a linear relationship between the buoyancy flux and the vertical buoyancy gradient~\citep{Chang:OM:05}

\begin{equation}
\frac{g}{\overline\rho}{\langle\rho'w'\rangle}=K_{\textrm{t}} N^2~,
\end{equation}

\noindent where $K_{\textrm{t}}$ is the eddy diffusivity and $N$ the buoyancy frequency.

%
%
%
Note that the choice of the relevant energy flux describing irreversible mixing is under debate \citep{tailleux_energetics_2009}. However, the eddy diffusivity defined above is directly measurable in the present configuration and provides a link between flow dynamics and irreversible turbulent fluxes. 

In addition, since the {TRI} process discussed in section~\ref{valid} has been suggested to participate in the energy cascade towards mixing scales~\citep{mackinnon2013parametric}, we will evaluate the quantity $K_{\textrm{t}}$ in cases with and without {TRI}. Indeed, the secondary waves fulfilling the resonance condition have smaller wavelengths than the forcing internal waves, facilitating the production of irreversible mixing. 

\begin{figure}
\includegraphics[width=0.48\textwidth]{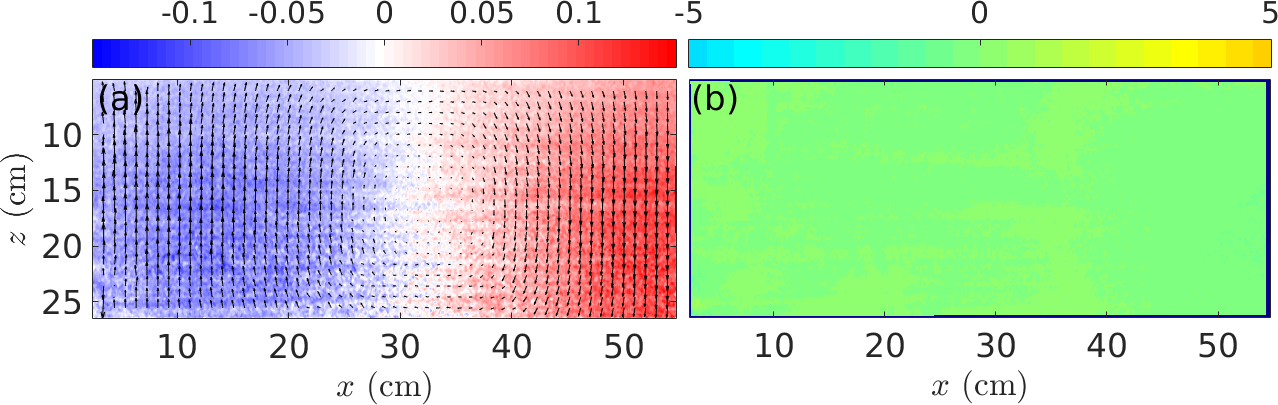} 
\caption{{\bf (a)} Density anomaly fields $\rho'(x,z,t=12~T_0)$ (in kg.m$^{-3}$) superimposed with velocity vectors in exp$A$. {\bf (b)} Associated turbulent diffusivity field $K_{\textrm{t}}(x,z,t=12~T_0)$ (in mm$^{2}\cdot$s$^{-1}$). Experimental parameters: $\omega_0/N=0.50$, mode-1 plate amplitude: $1$~cm.}
\label{fig:fluxnopsi}       
\end{figure}

Two experiments exp$A$ and exp$B$ combining simultaneous PLIF and PIV measurements are carried out to investigate the evolution of $K_{\textrm{t}}$. The mode-1 plate amplitude is $1$~cm in both experiments. In exp$A$, the nondimensional forcing frequency $\omega_A/N=0.50$ is too low for the experiment to exhibit any {TRI} instability. The radiation of a periodic mode-1 internal wave is observed, as shown in Fig.~\ref{fig:fluxnopsi}a. 
After two to three forcing periods, the interaction between the emitted and the reflected waves leads to a regime of stationary waves. No {TRI} disturbances are observed in the course of exp$A$.

\begin{figure}
\begin{center}
\includegraphics[width=0.4\textwidth]{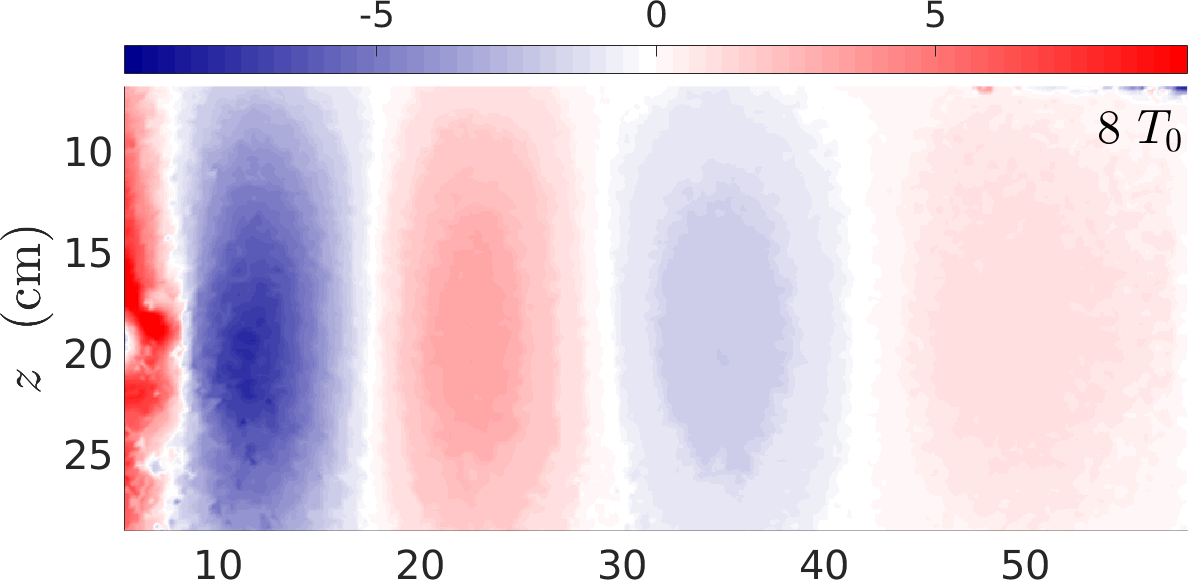} \\
\includegraphics[width=0.4\textwidth]{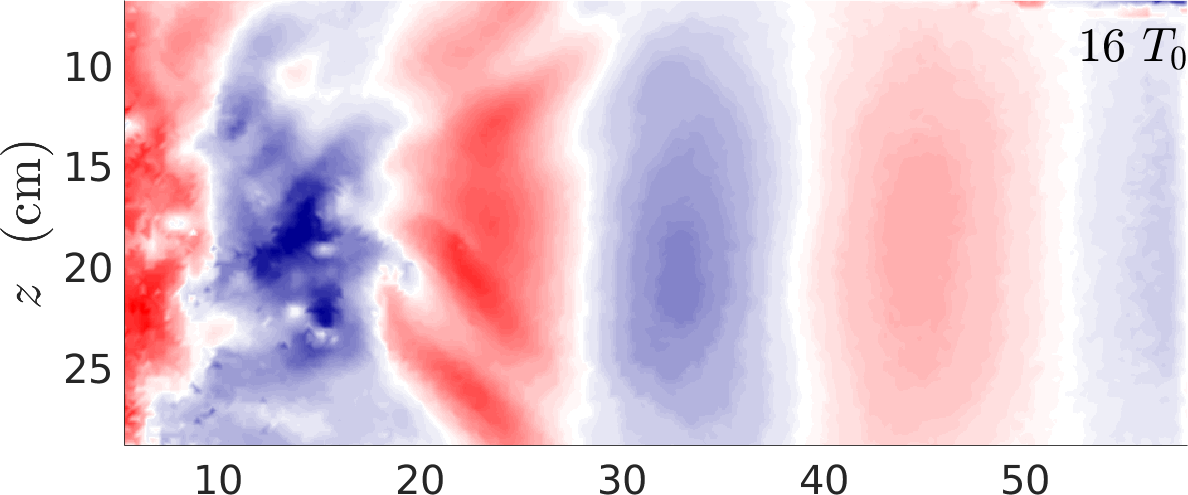} \\
\includegraphics[width=0.4\textwidth]{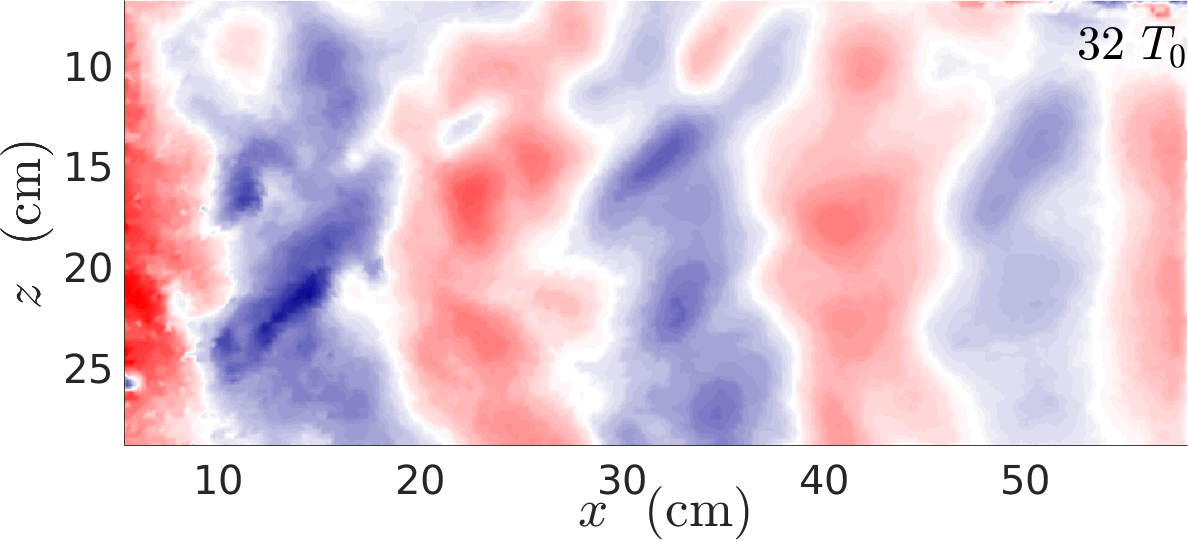} 
\caption{Vertical velocity field $w(x,z,t)$ (mm$\cdot$s$^{-1}$) issued from PIV measurements in exp$B$ at $t=8$~$T_0$ (top), $t=16$~$T_0$ (center) and $t=32$~$T_0$ (bottom).}
\label{fig:wpsi}       
\end{center}
\end{figure}

\begin{figure}
\includegraphics[width=0.48\textwidth]{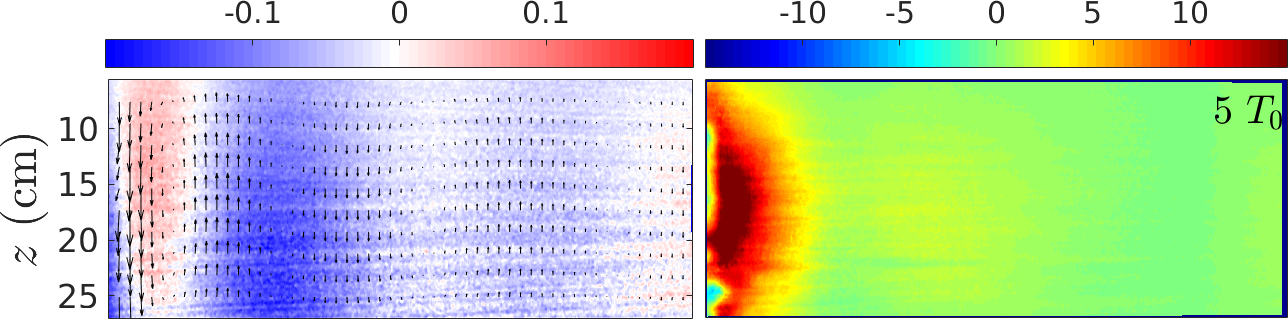} \\
\includegraphics[width=0.48\textwidth]{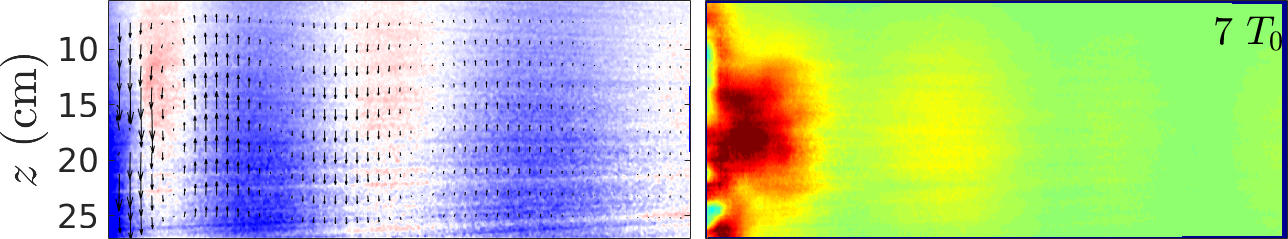} \\
\includegraphics[width=0.48\textwidth]{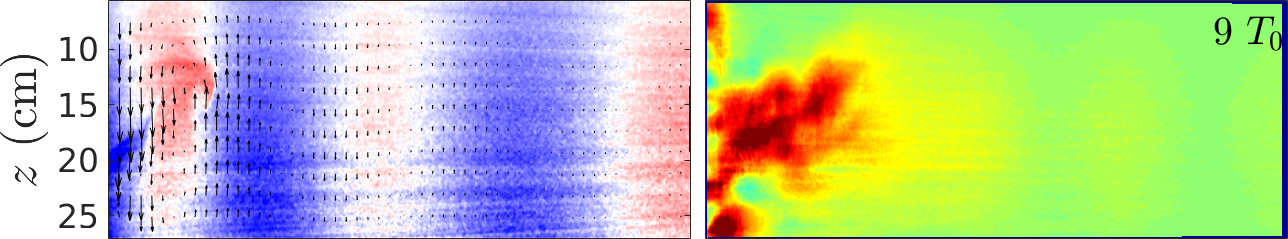} \\
\includegraphics[width=0.48\textwidth]{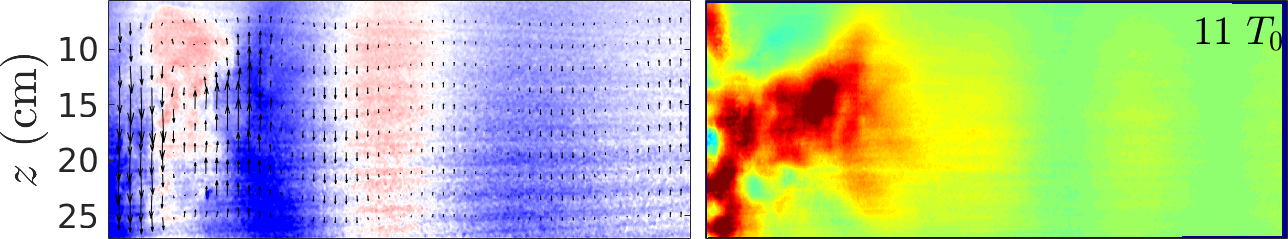} \\
\includegraphics[width=0.48\textwidth]{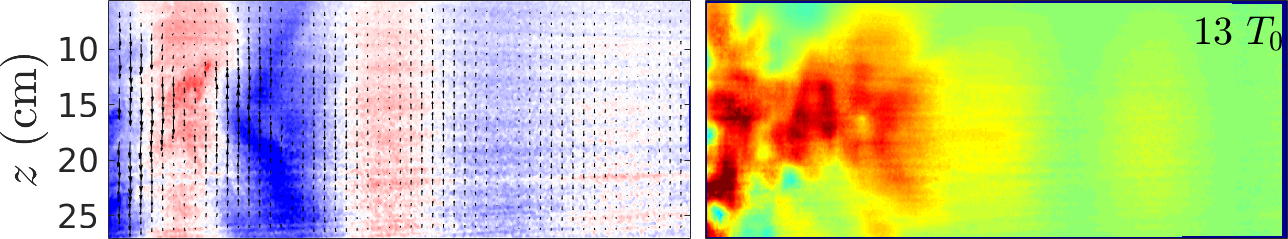} \\
\includegraphics[width=0.48\textwidth]{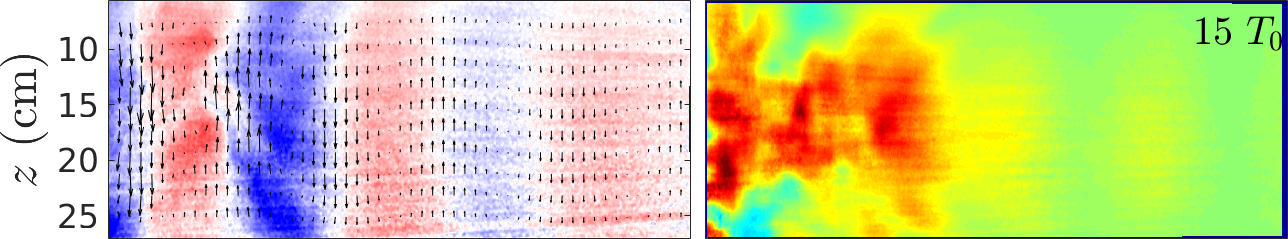} \\
\includegraphics[width=0.48\textwidth]{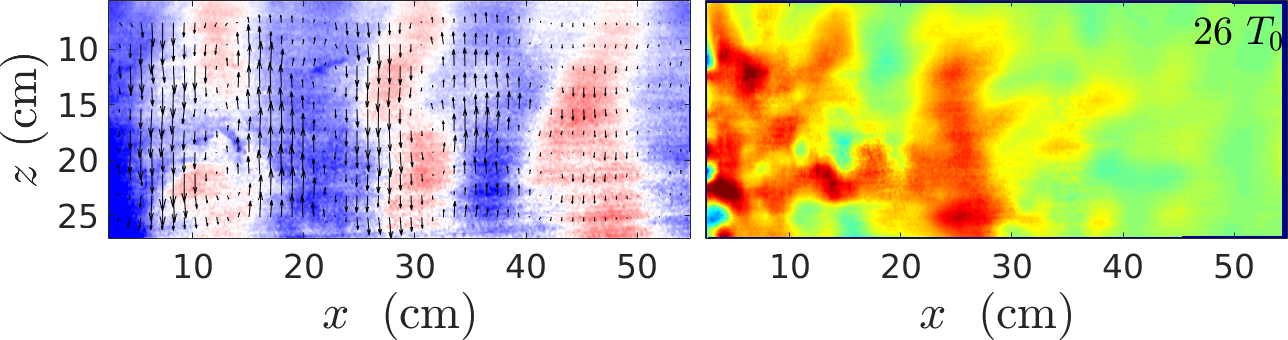} 
\caption{Left {panels}, from top to bottom: Density anomaly fields $\rho'(x,z,t)$ (in kg$\cdot$m$^{-3}$), superimposed with velocity vectors in exp$B$ {for $t=$ 5, 7, 9, 11, 13, 15 and 26~$T_0$}. Right {panel}: Associated turbulent diffusivity fields $K_{\textrm{t}}(x,z,t)$ (mm$^{2}\cdot$s$^{-1}$). Experimental parameters: $\omega_0/N=0.95$, mode-1 plate amplitude: $1$~cm.}
\label{fig:fluxpsi}       
\end{figure}

On the contrary, in exp$B$ ($\omega_B/N=0.95$), secondary internal waves issued from the {TRI} mechanism are expected to develop efficiently \citep{Bourget:JFM:13}. Fig.\,\ref{fig:wpsi} shows subsequent steps of the {TRI} onset on the velocity field. First, a mode-1 internal wave is radiated {from left to right} and, later, reflected at the {right} end wall. After a few forcing periods, ripples superimposed to the linearly propagating normal mode indicate that smaller scale waves are at play. The primary internal wave mode degenerates into a couple of secondary internal wave beams {fulfilling} the temporal resonance condition for {TRI}. Towards the end of the experiment they occupy the whole tank. Although PLIF measurements are overall less smooth than PIV measurements, the secondary waves are also observed in the density anomaly field in Fig.~\ref{fig:fluxpsi}, left panels, {for
$t=9$, 11, 13, 15 and 26~$T_0$}.  


In both experiments, the eddy diffusivity at time~$t$ is computed using the average value for $\rho'w'(x,z,t')$ between $t-2T_{0}$ and $t+2T_{0}$. The impact of the flow dynamics on irreversible energy fluxes shows large differences between exp$A$ and exp$B$. In the former case {(see Fig.~\ref{fig:fluxnopsi}b)}, random fluctuations of $K_{\textrm{t}}$ of the order of $1~{\rm mm}^2\cdot {\rm s}^{-1}$ are observed, defining the noise level of this measurement. Hence no quantitative positive turbulent buoyancy flux is measured in the configuration of low forcing frequency. 

In exp$B$, a steady patch of intense eddy diffusivity with a typical magnitude of $10$ to $15~{\rm mm}^2\cdot {\rm s}^{-1}$ is observed (see right panels of Fig. \ref{fig:fluxpsi}). The largest values of~$K_{\textrm{t}}$ in this patch are reached at mid-depth, where the horizontal velocity shear has the largest amplitude. For increasing time, the patch of intense mixing spreads towards the tank interior. The vertical extent of the mixing region increases linearly with the distance to the generator. This vertical spreading matches the propagation of secondary wave beams at constant angles away from their generation zone. After $30$~$T_0$, heterogeneous patches {with} large $K_{\textrm{t}}$ values are measured over the whole tank length. 

Hence, quantitative turbulent buoyancy fluxes are enhanced in the region of TRI. They result in values of $K_{\textrm{t}}$ at least one order of magnitude larger than the background value measured in the experiment with no TRI. One can conclude that the TRI process participates in the direct energy cascade towards irreversible mixing. Moreover, on the technical side, the combined PIV-PLIF procedure described here allows us to provide a quantitative assessment of the turbulent diffusivity field $K_{\textrm{t}}$.

\begin{figure}
\includegraphics[width=0.48\textwidth]{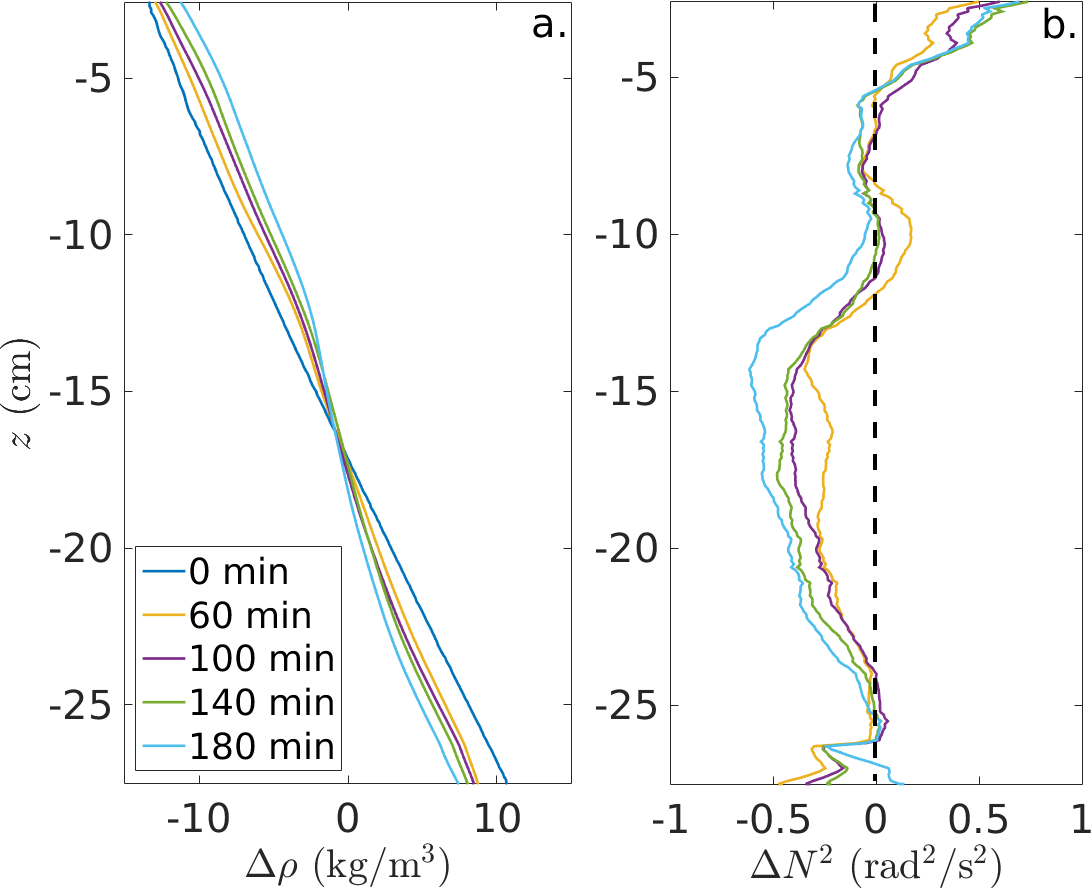} 
\caption{exp$C$: {\bf (a)} Background density profiles issued from PLIF images averaging. Note that the deformation of the profiles in the top 4 cm is due to local diffusion of the top mixed layer {\bf (b)} Squared background Brunt-Väisälä frequencies anomalies, obtained from derivating the curves in panel (a).}
\label{fig:evolstrat}       
\end{figure}

In order to assess the impact of the turbulent buoyancy fluxes on the irreversible mixing of the background stratification, a long term experiment exp$C$ was carried out over three hours of forcing, corresponding to 1800~$T_{0}$. The values for nondimensional forcing frequency and the forcing amplitude are chosen to be the same as in exp$B$. Conductivity probes allow for high resolution measurements of density profiles but this can be performed only when the fluid is at rest, since the conductivity probe cannot have a fast motion compared to flow velocities. However, it is also interesting to assess the evolution of the background stratification {\it in the course} of a mixing event. Averaging PLIF images recorded over one forcing period allows for the determination of the background stratification without stopping the forcing. Density profiles are obtained from PLIF images using a similar procedure as in section 2. Intensity profiles are averaged over 20 pixels closest to the tank end-wall, in order to minimize attenuation. Density profiles, displayed in Fig.~\ref{fig:evolstrat}a, are obtained after a calibration against a conductivity probe measurement before starting the experiment. Subsequent density profiles are overall smooth and capture the evolution of the background stratification. At $t=0$, the initial profile is linear ($N=1$ rad$\cdot {\rm s}^{-1}$). With increasing time, the background stratification progressively decreases in the center of the density profile while it remains sensibly constant in the top and bottom thirds of the profile. As the fluid becomes less stratified, the largest and weakest density values get closer to the value at the center. Irreversible mixing of the background stratification is also reflected in terms of changes in the squared Brunt-Väisälä frequency $N^2$ (figure~\ref{fig:evolstrat}b), obtained by derivating the density profiles. While values of $N^2$ remain close to the initial stratification in the top and bottom parts, a large decrease of $N^2$ down to $0.65$ rad$^2\cdot$s$^{-2}$ is measured in the central region, where the TRI process occurs. Turbulent diffusivities enhanced by the TRI process lead to substantial changes in the background stratification of the flow.  

\section{Conclusion}
\label{conclu}
The PLIF technique, which we describe and validate for a continuously stratified fluid, permits to carry out quantitative field density measurements. The internal wave dynamics is accurately captured using PLIF in a configuration of secondary internal wave radiation via the {TRI} process. 

Combining this technique with simultaneous and spatially synchronized velocity measurements via PIV allows for the measurement of correlations between density and velocity fields, and in particular gives access to the eddy diffusivity field $K_{\textrm{t}}$. The energy cascade occurring during the {TRI} process yields values of $K_{\textrm{t}}$ up to $15$~mm$^2$ $\cdot$ s$^{-1}$, which are {one} order of magnitude larger than the background eddy diffusivity for a linear normal mode propagation. The local mixing evidenced by this measurement converts, during larger duration experiments, into a deformation of the background density profile, highlighting the occurence of irreversible mixing, attesting a gain of potential energy, taken from the internal wave flow kinetic energy. This mixing takes place principally in the middle part of the tank, where the shear of the mode-1 is dominant.

The present values of $K_{\textrm{t}}$ are obtained in an idealized laboratory scale configuration and cannot be directly used to assess oceanic turbulent diffusivities. However, it is clear that {TRI} can substantially enhance internal wave induced mixing away from topographies for quasi-linear stratifications. Future experimental studies will involve a full parametric study of the system, and could investigate upon the relation between the dynamics of {TRI} and induced mixing for non-uniform background stratifications. 

\bibliographystyle{ExpFluids}       


%
%

\end{document}